\documentclass[preprint,12pt]{elsarticle}

\usepackage{amssymb}
\usepackage{amsthm}
\usepackage{amsmath}
\usepackage[colorlinks]{hyperref}

\journal{Physics of the Dark Universe}

\begin{document}

\begin{frontmatter}



\title{Searching for galactic axions through magnetized media: the QUAX proposal}

\author[label0]{R. Barbieri}
\author[label2]{C. Braggio}
\author[label2]{G. Carugno}
\author[label2]{C.S. Gallo}
\author[label1]{A. Lombardi}
\author[label1]{A. Ortolan}
\author[label1]{R. Pengo}
\author[label1]{G. Ruoso}
\ead{Giuseppe.Ruoso@lnl.infn.it}
\author[label3]{C.C. Speake}
\address[label0]{Institute of Theoretical Studies, ETH, CH-8092 Zurich (Switzerland) and Scuola Normale Superiore, 56100 Pisa (Italy)}
\address[label2]{INFN, Sezione di Padova and Dipartimento di Fisica e Astronomia, Via Marzolo 8, 35131 Padova  (Italy)}
\address[label1]{INFN, Laboratori Nazionali di Legnaro, Viale dell'Universit\`a 2, 35020 Legnaro (Italy)}
\address[label3]{School of Physics and Astronomy, University of Birmingham, West Midlands B15 2TT (UK)}

\begin{abstract}
We present  a proposal to search for QCD axions with mass in the 200 $\mu$eV range, assuming that they make a dominant component of dark matter. Due to the axion-electron spin coupling, their  effect is equivalent to the application of an oscillating rf field with frequency and amplitude   fixed by the axion mass and coupling  respectively. This equivalent magnetic field would produce spin flips in a magnetic sample placed inside a static magnetic field, which determines the resonant interaction at the Larmor frequency. Spin flips would subsequently emit radio frequency photons that can be detected by a suitable quantum counter in an ultra-cryogenic environment. This new detection technique is crucial to keep under control the thermal photon background which would otherwise produce a 
too large  noise.
\end{abstract}

\end{frontmatter}



\section{Introduction}
An outstanding result of  modern cosmology is that a significant fraction of the universe is made of dark matter. However, the nature of such component is still unknown, apart its gravitational interaction with ordinary baryonic matter. A favored candidate for dark matter is the axion: a new particle introduced by Peccei and Quinn   to solve the strong CP problem \cite{PQ}, i.e. the absence of CP violation in the strong interaction sector of the Standard Model. Axions have a mass $m_a$ inversely proportional to the Peccei-Quinn  symmetry breaking scale $f_a$.
For certain ranges of $f_a$ and $m_a$ (typically with masses ranging from $\mu$eV to meV), large quantities of axions may have been produced in the early Universe that could  account for a  portion or even the totality of  cold dark matter. Axions have extremely small coupling to normal matter and radiation, but they can be converted into detectable photons by means of the inverse Primakoff effect as shown by Sikivie \cite{Sikivie}. The idea of Sikivie has been exploited by  several experiments \cite{MelissinosPRL, MelissinosAX, Florida}, of which the most recent  is ADMX \cite{ADMX,ADMX2}. The latter experiment is still running, and for the moment it has been capable of exploring the axion model for masses of a few $\mu$eV \cite{ADMX3}. 

The QUAX (QUaerere AXion) proposal explores in details the ideas of Ref.s \cite{Krauss1985, Barbieri1989,Yannis,Kolo1991,Voro1995}.  These authors proposed to study the interaction of the cosmological axion with the spin of fermions (electrons or nucleons). In fact, due to the motion of the Solar System through the galactic halo, the Earth is effectively moving through the cold dark matter cloud surrounding the Galaxy and an observer on Earth will see such axions as a wind. In particular, the effect of the axion wind on a magnetized material can be described as an effective oscillating rf  field with frequency determined by $m_a$ and amplitude related to $f_a$. Thus, a possible detector for the axion wind can be a magnetized sample with Larmor resonance frequency tuned to the axion mass by means of an external polarizing static magnetic field: e.g. 1.7 T for 48 GHz, corresponding to a 200 $\mu$eV axion mass, in the case of the interaction with the electron spin that is considered hereafter. The interaction with the axion effective field will drive the total magnetization of the sample, and so produce oscillations in the magnetization that, in principle, can be detected.
In order to optimize the detection scheme, the sample is placed inside a microwave cavity. The cavity and the magnetized sample have to be cooled down at ultra - cryogenic temperature to avoid the noise due to  thermal photons. 

Within all axion models \cite{Kim}, this detection scheme is sensitive only to DFSZ axions \cite{DFSZ,DFSZ0,DFSZ1}. For example, in the KSVZ  model \cite{KSVZ,KSVZ1} the electron coupling is strongly suppressed.

The paper is organized as follows. For ease of the reader we give in Section 2 an introduction to the calculation of the effective magnetic field due to the axion wind. 
After that the experimental scheme will be presented in Section 3 and its sensitivity calculated in Section 4. 
A crucial role is played by the detector of the magnetization changes: two different approaches are discussed, namely a linear rf amplifier and a microwave quantum counter. We will show that the latter approach can allow to investigate the parameter space of DM axions. However,  preliminary measurements with linear amplifiers could set limits on the axion-electron coupling constant close to the expected values, as discussed in Sections 4 and 5.

The QUAX R\&D activities are conducted at the Laboratori Nazionali di Legnaro (LNL) of the Istituto Nazionale di Fisica Nucleare (INFN), 
and funded in the framework of the research call {\it What Next} of INFN.

\section{The axion wind and the effective magnetic field}

For ease of the reader we recall the basic elements that allow to calculate the value of the effective magnetic field due to the presence of the axion wind. The  Lagrangian which describes the interaction of a spin 1/2 particle with the axion field $a(x)$    reads
\begin{equation}\label{eq1}
L=\bar{\psi}(x)(i\hbar 
\gamma^\mu\partial_\mu- mc)\psi(x) - ig_p a(x) \bar{\psi}(x)\gamma_5\psi(x),
\end{equation}
where $\psi(x)$  is the spinor field of the fermion with mass $m$. Here $\gamma^\mu$ are the 4 Dirac matrices, $\gamma^5=i \gamma^0\gamma^1\gamma^2\gamma^3$, and $a(x)$  is coupled to matter by the  dimensionless pseudo-scalar coupling constant $g_p$. By taking the non-relativistic limit of the Euler-Lagrange equation,  the time evolution of a  spin 1/2 particle can be described by the usual Schroedinger equation
\begin{equation}\label{eqsh}
i\hbar\frac{\partial \varphi}{\partial t}=\left[-\frac{\hbar^2}{2m}\nabla^2 - \frac{g_p\hbar }{2m}\boldsymbol{\sigma}\cdot \boldsymbol{\nabla}a \right]\varphi\ , 
\end{equation}
where the  term 
\begin{equation}
 -\frac{g_p\hbar }{2m}\boldsymbol{\sigma}\cdot \boldsymbol{\nabla}a \equiv -2 \frac{e \hbar }{2m}\boldsymbol{\sigma}\cdot\left(\frac{g_p}{2e}  \right)\boldsymbol{\nabla}a
\label{spin_int}
\end{equation} 
has the form of the interaction between the spin magnetic moment ($-2 \frac{e \hbar }{2m}\boldsymbol{\sigma} = - 2 \mu_B \boldsymbol{\sigma}$, with $\mu_B$ the Bohr magneton in the case of the electron) and an effective 
 magnetic field $B_a \equiv \frac{g_p}{2e}  \boldsymbol{\nabla}a $ . 
As in the original Ref.s \cite{ Barbieri1989,Kolo1991,Voro1995} we do not consider here the parity-violating term $-i \frac{m_a}{m}\frac{a}{f_a}\boldsymbol{\sigma}\cdot \boldsymbol{p}$, which induces  oscillating electric dipole moments in atoms \cite{Stadnik:2013raa,Graham:2013gfa}. The interaction (\ref{spin_int}) has been recently considered in Ref.s \cite{Sikivie:2014lha, Axioma} to search for axion induced atomic transitions using laser techniques.

Axions represent the best example of non-thermal dark matter candidate \cite{Turner}. 
The expected dark matter density is $\rho\simeq 300$ MeV/cm$^3$, and we will suppose that axions are the dominant component.  For an axion mass  in the range
 $ 10^{-6}{\rm eV} < m_a < 10^{-2} {\rm eV}$, we have $n_a\sim 3\times10^{12}\  (10^{-4}\ {\rm eV}/m_a)$ axions per cubic centimeter. The axion velocities $\boldsymbol{ v}$ are distributed in modulus according to a Maxwellian distribution,
 with a velocity dispersion
$\sigma_v\approx 270$ km/sec.
Due to Galaxy rotation and Earth motions in the Solar system, the rest frame of an Earth based laboratory is moving through the local axion cloud with a time varying velocity $\boldsymbol{v}_E= \boldsymbol{ v}_S+ \boldsymbol{ v}_O+ \boldsymbol{ v}_R$, where  $\boldsymbol{ v}_S$  represents the Sun velocity in the galactic rest frame (magnitude 230 km/sec), $\boldsymbol{ v}_O$  is the Earth's orbital velocity around the Sun (magnitude 29.8 km/sec), and $ \boldsymbol{ v}_R$ the Earth's rotational velocity (magnitude 0.46 km/sec).  
The observed axion velocity is then $\boldsymbol{ v}_a = \boldsymbol{ v}- \boldsymbol{ v}_E$. The effect of this motion is to broaden the Maxwell distribution, as well as to modulate it with a periodicity of one sidereal day and one sidereal year. 

The axion kinetic energy  is expected to be distributed with a mean relative to the rest mass of  $7 \times 10^{-7}$ and a dispersion about the mean of $5.2 \times 10^{-7}$ \cite{turner1}. The inverse of this last number represents the natural figure of merit of the axion linewidth, $Q_a \simeq 1.9\times 10^6$.
The mean De Broglie wavelength of an axion  is (we will use in the following as central reference value in our calculations the axion mass $m_a = 200 \,\mu$eV)
\begin{equation}
\lambda_d\simeq {h}/ ({m_a v_a}) \simeq  6.9 \left(\frac{200\, \mu{\rm eV}}{m_a}\right) \,\,\, {\rm m},
\end{equation}
 which is much greater than the typical length of an experimental apparatus, in our case the magnetized samples. 

 Such theoretical and experimental aspects allow  to  treat $a(x)$ as a classical field that interacts coherently with  fermions with a mean value 
$a(x)=a_0 {\rm exp}[ {i ( p^0 c t-\boldsymbol{p}_E\cdot\boldsymbol{x} )/\hbar}]$
where $\boldsymbol{ p}_E = m_a \boldsymbol{ v}_E$, $c p^0=\sqrt{m_a^2 c^4 + |\boldsymbol{ p}_E |^2 c^2} \approx m_a c^2 + |\boldsymbol{p}_E|^2 /(2 m_a)$ and $a_0$ is the field amplitude.
The amplitude $a_0$ can be easily computed by equating the momentum carried by this field per unit  volume (i.e. the associated energy momentum tensor $T^{0i}=a_0^2 p^0 p_E^i$) to the number of axion per unit volume times the average momentum (i.e. $n_a<p^i> =n_a p_E^i$), and it reads 
$a_0=\sqrt{({n_a\hbar^3 })/({m_a c}})$.
The effective magnetic field associated with the mean axion field is then given by
 \begin{equation}
\mathbf{B}_a = \frac{g_p}{2 e}  \left(\frac{n_a\hbar }{m_a c}\right)^{1/2}\, \boldsymbol{ p}_E \,  \sin\left(\frac{p^0 c t-\boldsymbol{p}_E\cdot\boldsymbol{x}}{\hbar}\right).
\end{equation} 
{\color{black}
In the framework of DFSZ  axion model \cite{DFSZ,DFSZ0,DFSZ1}, the value of the coupling constant $g_p$ with electrons  can be expressed as $g_p = m_e/ (3 f_a)\cos^2 \beta$, where $\cos^2 \beta$ is a model dependent parameter, here fixed to 1,  $f_a$ is the axion decay constant \cite{rpp,Grilli} and
\begin{equation}
    m_a \simeq 5.7  \, \mu{\rm eV} \left(\frac{10^{12} \, {\rm GeV}}{f_a} \right).
\end{equation}
Thus $g_p \simeq  3.0 \times 10^{-11} ({m_a}/({1\  \rm{eV}}))  $.  
}

Putting the magnetized samples in $\boldsymbol{ x} =\boldsymbol{ 0}$, we have that the equivalent oscillating rf field along the $\boldsymbol{ p}_E$ direction has a mean amplitude and central frequency 
\begin{equation}
B_a  = 2.0 \cdot 10^{-22} \left(\frac{m_a}{200\, \mu {\rm eV}}\right)    \,\,\,\, {\rm T},\,\,\,\,\, \frac{\omega_a}{2 \pi}  =  48  \left(\frac{m_a}{200\,\mu{\rm eV}}\right)    \,\,\,\, {\rm GHz}, \, 
\label{axionfield}
\end{equation}
with a relative linewidth $\Delta \omega_a/\omega_a \simeq 5.2 \times 10^{-7}$.  

{\color{black}
As the equivalent magnetic field  is not directly associated to the axion field but to its gradient, the corresponding coherence time  and correlation length  for the QUAX detector read  
\begin{eqnarray}
\tau_{\nabla a} &\simeq& 0.68\, \tau_{a} =  17\left(\frac{200\, \mu{\rm eV}}{m_a}\right) \left( \frac{Q_a}{1.9 \times10^6}\right)\,\,\,\,\mu{\rm s};  \nonumber \\  
\lambda_{\nabla a} &\simeq& 0.74\, \lambda_a = 5.1 \left(\frac{200\, \mu{\rm eV}}{m_a}\right)\,\,{\rm m,}\label{coherence}
\end{eqnarray}
 where we have assumed  the standard CDM halo model \cite{turner1} for the axion velocity distribution. 
}

\section{Experimental scheme}

To detect the extremely small rf field $B_a$ we will make use of the Electron Spin Resonance (ESR) in a magnetic sample. In particular, we want to collect the power deposited in the sample by the axion wind due to its interaction with the electron spin. To enhance the interaction we will tune the ferromagnetic resonance of the sample, i.e. the Larmor frequency of the electron, to the mass value of the searched for axion. In fact, since this is still unknown, the possibility to perform a large bandwidth search must be envisaged.

Let us consider a magnetized  sample of volume $V_s$ and magnetization $M_0$  placed  in the bore of a solenoid, which generates a static magnetic field $B_0$ (polarizing field). 
The value $B_0$  determines the Larmor frequency of the electrons, 
and so the axion mass  under scrutiny, through the relation  $(\gamma = e/m_e)$
\begin{equation}
B_0 = \frac{\omega_L} {\gamma} = \frac{ m_a c^2}{\gamma \hbar } = 1.7 \left(\frac{m_a}{200\, \mu {\rm eV}}\right)  \,\,\,\, {\rm T}.
\end{equation}
The dynamics of the magnetic sample is well described by its magnetization $\mathbf{ M}$, whose evolution is given by the  Bloch equations with dissipations  and radiation damping \cite{bloom}. Taking the external magnetic field directed along the $z$ axis, one has
\begin{eqnarray}\label{bloch}
\frac{dM_x}{dt}&=&\gamma  (\mathbf{M}\times\mathbf{ B})_x - \frac{M_x}{\tau_2} -\frac{M_x M_z}{M_0 \tau_r} \nonumber \\
\frac{dM_y}{dt}&=&\gamma (\mathbf{M}\times\mathbf{ B})_y - \frac{M_y}{\tau_2} - \frac{M_y M_z}{M_0\tau_r} \nonumber \\
\frac{dM_z}{dt}&=&\gamma (\mathbf{M}\times\mathbf{ B})_z - \frac{M_0- M_z}{\tau_1}- \frac{M_x^2+M_y^2}{M_0 \tau_r},
\end{eqnarray}
where $M_0$ is the static magnetization directed along the $z$ axis\footnote{This is true for paramagnets. In general for other type of samples a suitable  orientation of the axes is necessary.}, $\tau_r$ is the radiation damping time, $\tau_1$ and $\tau_2$ are the longitudinal (or spin-lattice) and transverse (or spin-spin) relaxation time, respectively. We will discuss the role of radiation damping in a subsequent paragraph. Moreover, we will assume that $\tau_1$ is never shorter than $\tau_2$, which is true for most of the materials we are interested in.

In the presence of the axion effective field $B_a (t)$ a time dependent component of the magnetization $ {M}_a(t)$  will appear in the $x-y$ plane. We note here that only the component of the axion effective field orthogonal to the magnetizing field $B_0$ will drive the magnetization of the sample, thus configuring this apparatus as a true directional detector. It is
\begin{equation}
M_a (t)= \gamma \mu_B B_a n_S \tau_{\rm min} \cos(\omega_a t),
\label{axmag}
\end{equation}
where $n_S$ is the material spin density and $\tau_{\rm min}$ is the shortest coherence time among the following processes: axion wind coherence $\tau_{\nabla a}$, magnetic material relaxation $\tau_2$, radiation damping $\tau_r$
\begin{equation}
\tau_{\rm min}=\min\left(\tau_{\nabla a},\tau_2,\tau_r\right).
\end{equation}

The axion coherence time $\tau_{\nabla a}$ has been given in Equation (\ref{coherence}).
Material relaxation times span normally from tens of nanoseconds up to a few microseconds.

The radiation damping $\tau_r$ was introduced by Bloom in 1957 \cite{bloom} to account for the electromotive force  induced in the rf coils of a driving circuit by magnetization changes without 
taking into account the dynamics of the rf coils. 
In our case this damping term is affecting the maximum allowed coherence hence the integration time of the magnetic system with respect to the axion driving input. In a free field environment and in a   high  frequency regime (GHz or more), which is our case, radiation damping is dominated by the magnetic dipole  emission in free space from the magnetized sample of volume $V_s$
\begin{equation}
\tau_r=4 \pi  \frac{c^3}{\omega_L^3} \frac{1}{ \gamma \mu_0 M_0 V_s } \label{dampingtime}.
\end{equation}
As already mentioned, we are considering a situation in which the coherence lenght of the axion field is much larger than the typical size of the magnetized sample.

The steady state solutions of  Equation (\ref{bloch}) in presence of radiation damping and for various approximations  are given in ref. \cite{augustine}.  
Since we are working at high frequency, the radiation damping mechanism may result in a strong limitation. To solve this problem we embed the magnetic material inside a microwave resonant cavity in the strong coupling regime. In this case, the limited phase space of the resonator inhibits the damping mechanism, thus providing a minimum radiation damping time equal to the cavity decay time. Now we have $\tau_{\rm min}=\min\left(\tau_{\nabla a},\tau_2, \tau_c \right)$, where $\tau_c$ is the cavity decay time.

\begin{figure}[htbp]
\begin{center}
\includegraphics[width=.75\linewidth]{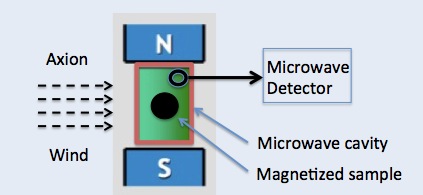}
\caption{Principle scheme of the axion haloscope}
\label{schema}
\end{center}
\end{figure}

Figure \ref{schema} shows the principle of the proposed detection scheme: a microwave resonant cavity, containing a magnetic material, is kept at very low temperature and placed inside an extremely uniform magnetizing field $B_0$. 
The magnetic field  value determines the Larmor frequency $\nu_m= \gamma B_0 /2 \pi$ of the ferromagnetic resonance of the magnetic material. A resonant mode of the microwave cavity, typically the TE120,  is tuned to  the Larmor frequency, in such a way that the electron spin couples to the electromagnetic field stored in the cavity. The single spin coupling is $g_0 = \gamma \sqrt{\mu_0 \hbar \omega_m/ V_c}$ (In  units of rad/s) \cite{Tabuchi2014}, where $V_c$ is the volume of the cavity mode. If the total number of spin is large enough, hybridization takes place and the single resonance splits into two with a mode separation given by the total coupling strength $g_m = g_0 \sqrt{n_S V_s}$. The transmission coefficient of the hybridized system is now described by
\begin{equation}
S_{21}(\omega)\simeq \frac{1}{i (\omega - \omega_c) -\frac{k_c}{2}+ 
\frac{|g_m|^2}{i(\omega - \omega_m)-k_m/2}} \label{hyb},
\end{equation}
where $k_c = 1/\tau_c$ is the total linewidth of the cavity with resonance frequency $\omega_c$ and
$k_m=1/\tau_2$ is the linewidth of the ferromagnetic resonance. The resulting linewidth $k_h$ of a single hybridized mode is
\begin{equation}
k_h =  \frac{1}{2}\left(k_c + k_m\right).
\end{equation}
When the system is hybridized the dynamics is completely described by the hybrid mode characteristic time $\tau_h = 1 / k_h$. Radiation damping is not  effective anymore. In the following we will also assume that the material spin - lattice relaxation time $\tau_1$ is greater than $\tau_h$, so that now $\tau_{\rm min}=\min\left(\tau_{\nabla a},\tau_h \right)$.

In the presence of the axion wind, 
the average amount of power absorbed by the material in each cycle is
\begin{eqnarray}
P_{\rm in} &=& \mu_0 {\bf H} \cdot \frac{d{\bf M}}{dt} =B_a \frac{d M_a}{dt}  V_s \nonumber \\
&=& \gamma \mu_B n_S  \omega_a B_a^2   \tau_{\rm min} V_s \label{powerin},
\end{eqnarray}
 where we have used Equation (\ref{axmag}) for the axion induced resonant magnetization $M_a$. 
 In a steady state, the power balance ensures that $P_{\rm in}$ will be emitted as rf radiation, and so $P_{\rm in}/2$ can  be collected by using an antenna critically coupled to the cavity mode. 

 It is useful to write the output power by referring to relevant experimental design parameters
\begin{equation}
P_{\rm out}=\frac{P_{\rm in}}{2}=
3.8 \times 10^{-26} \left(\frac{m_a}{200\, \mu{\rm eV}}\right)^3 \left( \frac{V_s}{100\,\, {\rm cm}^3}\right)
 \left( \frac{n_S}{2 \cdot10^{28} /{\rm m}^3}\right)
 \left( \frac{\tau_{\rm min}}{2 \,\mu{\rm s}}\right)\,{\rm W},
 \label{power}
\end{equation}
where the chosen axion mass is determined by a magnetizing field  $B_0=1.7$ T, and the value of the spin density is typical of paramagnets at low temperature or  material as YIG (Yttrium Iron Garnet) \cite{YIG} also at room temperature. A geometrical factor of the order of 1, which takes into account  the coupling between the material magnetization  and cavity modes, has been neglected.

Measuring such a low power is a  difficult task. Following a discussion by Lamoreaux \cite{Lamoreaux2013} a single photon detection would be the best choice. The expected rate of emitted photons is
\begin{equation}
R_a = \frac{P_{\rm out}}{\hbar \omega_a}= 1.2 \times 10^{-3} \left(\frac{m_a}{200\, \mu{\rm eV}}\right)^2 \left( \frac{V_s}{100\,\, {\rm cm}^3}\right)
  \left( \frac{n_S}{2 \cdot10^{28} /{\rm m}^3}\right)
 \left( \frac{\tau_{\rm min}}{2 \,\mu{\rm s}}\right)\,{\rm Hz} \label{rateaxion}.
\end{equation}

A similar calculation can be done by using the single spin flip transition probability, which is given by $ c_1 \gamma^2 B_a^2 \tau $, where $c_1$ is a coupling coefficient of order one and $\tau$ is the transition relaxation time. By taking $\tau=\tau_{\rm min}$ and $c_1=1$ one finds for the emitted photons:
\begin{equation}\label{axionrate}
R_a^* =\frac{1}{2}  \gamma^2 B_a^2 \tau n_S V_s =
\end{equation}
\begin{equation}
= 2.4 \times 10^{-3}\left(\frac{m_a}{200\, \mu{\rm eV}}\right)^2 \left( \frac{V_s}{100\,\, {\rm cm}^3}\right)
  \left( \frac{n_S}{2 \cdot10^{28} /{\rm m}^3}\right)
 \left( \frac{\tau_{\rm min}}{2 \,\mu{\rm s}}\right)\,{\rm Hz} \nonumber.
  \end{equation}
  
{\color{black} 
It is interesting to compare 
Equation (\ref{rateaxion}) with the photon rate associated to the axion - photon conversion through the Primakoff effect \cite{Sikivie}, as  envisaged in the ADMX experiment \cite{ADMX}. From Eq (3.9) of ref. \cite{Sikivie1985} we have 
\begin{equation}
R_a^{\gamma}\simeq 10.0 \times 10^{-3}\,\,  C_{nl} \left(\frac{g_{\gamma}}{0.36}\right)^2  \left( \frac{V_s}{100\,\, {\rm cm}^3}\right)
 \left( \frac{B_0}{{\rm 2\, T}}\right)^2
 \left( \frac{\tau_{\rm min}}{2 \,\mu{\rm s}}\right)\,{\rm Hz}\, , \label{primakoff}
 \end{equation}
where $C_{nl}$ is a factor of the order of 1 describing the overlap between the radiofrequency electric field of the cavity resonant mode ${nl}$ and the external magnetizing field $B_0$;  here $g_\gamma$ is an axion parameter which is equal to  0.36 or 0.97 in the DFSZ or KSVZ models, respectively. It is worth noticing that the expected photon rate via Primakoff effect can be larger than the rate due to the interaction with the electron spins. However, with a suitable choice of cavity modes coupled to the photon counter (e.g. TE110 or TE120 mode of a rectangular cavity), the  QUAX apparatus can detect separately Primakoff and spin flip photons, allowing us to distinguish between DFSZ and KSVZ axions.
This paper is focused on the detection of axion electron coupling in a polarized magnetic material. However,  QUAX cavities can be also operated in the Sikivie  configuration, thereby increasing the overall axion detection capability.  We stress that both DFSZ and KSVZ axion  searches are important, as the QCD axion model  has not yet been validated.

}

\section{Expected sensitivity}

\subsection{Thermal noise in a single photon counter}

As shown below it is hard if not impossible to measure
the power calculated in Equation (\ref{power})  using a linear amplifier  in a reasonable amount of time. 
As reported in Ref.  \cite{Lamoreaux2013}, the possibility of using  a single photon counter can overcome the problem. In this work by Lamoreaux  an extensive discussion on the possibility of developing a single photon counter in the microwave regime is given. 
More recently, a proposal has been published where a 30 GHz single photon counter is envisaged \cite{Stax}. In such a paper a suitably designed superconducting antenna is connected to a nano-sized hot-electron calorimeter read out by an ultra low noise superconducting quantum interference device (SQUID) amplifier. Such system could provide a single photon detector whose limit is the thermal photon bath.
In view of these developments we discuss the expected sensitivity assuming that the single photon detection is available. 

In the absence of technical noise, the ultimate noise source is the thermal bath. The average number of photon $\bar{n}$ in an empty cavity at temperature $T_c$ is given by
\begin{equation}
\bar{n} = \frac{1}{e^\frac{\hbar \omega_c}{k_B  T_c}-1}. \label{avphoton}
\end{equation}
Therefore the average energy stored in the cavity is $E_c = (\bar{n} +1/2) \hbar \omega_c$,  
whereas the rate of the emitted photons is
\begin{equation}
R_t = \bar{n}/\tau_c \label{rate}.
\end{equation}

For a cavity filled with a magnetic material we expect that the rate is still given by Equation (\ref{rate}), since the power of fluctuations of cavity and material cannot exceed the power injected by the thermal bath. See Section 5 for a first attempt at an experimental verification of this statement in a preliminary test. 

Suppose now that we are measuring with a technical noise-free quantum counter: after a measurement time $t_m$ the total number $N$ of photons detected (axion induced + thermal) will be 
\begin{equation}
N = \eta (R_a + R_t) t_m,
\end{equation} 
where $\eta$ is the quantum efficiency of the detector. 
The total signal to noise ratio will be given by
\begin{equation}
{\rm SNR} = \frac{\eta R_a  t_m}{\sqrt{\eta (R_a + R_t) t_m}} =\frac{R_a}{\sqrt{R_a+R_t}}\sqrt{\eta t_m}.
\end{equation}

In order to choose an appropriate measurement time, we have to verify that the intrinsic noise is low enough. For the rate given by Equation (\ref{rateaxion}), by setting $t_m = 1.4\times 10^4$~s (about four hours) we would get on average 17 counts, i.e. an intrinsic  signal to noise ratio of about $17/\sqrt{17} \simeq 4$, which is reasonable. Taking into account the thermal background we can have  the usual request that a signal  can be detected at ${\rm SNR} =3$. This will limit the rate of the thermal photon counting, i.e. the temperature of the device. Assuming $\eta \sim 1$  the maximum rate of thermal photons must be
\begin{equation}\label{ratefinal}
R_t = R_a \left( \frac{R_a \eta t_m}{{\rm SNR}^2}-1 \right)  \sim 1.0 \times 10^{-3}\,{\rm Hz}.
\end{equation}
 
By means of  equations (\ref{avphoton}) and (\ref{rate}), 
this condition  requires a  maximum temperature of
116 mK at the working frequency of 48 GHz and with cavity decay time of 2 $\mu$s.

The bandwidth of the measurement is given by $2 k_h/2 \pi \sim 150$ kHz, i.e. the total  width of the two hybridized modes in the cavity-material system. In a month of operation about 100  measurements can be performed at  different central frequencies, for a total span of about 15 MHz. Frequency tuning can be accomplished by using sapphire tuners, while only a modest change in the magnetic field is necessary. Over longer time scales a wider bandwidth can be explored.

Needless to say,  at this level of background other sources of energy should be taken into account, such as cosmic ray radiation and natural radioactivity. Suitable shielding of the apparatus should be planned, together with the choice of low background materials.

\subsection{The maximum sensitivity of a linear amplifier}

To evaluate the maximum sensitivity that could be reached with a linear amplifier we note that an axion signal will manifest itself as a peak with width $\Delta f = \Delta \omega_a/(2\pi)$. This corresponds to a maximum of two bins   in the output spectrum if one chooses as resolution bandwidth ${\rm RBW} = \Delta f $. Two bins is the worst situation where the axion signal is divided exactly into two adjacent bins. By power averaging the detected spectrum it is possible to reduce the standard deviation of the expected power deposited in a single bin, so as to better identify the signal, i.e. a peak summed up to the Lorentzian shape of the cavity thermal noise. This is equivalent to say that the minimum measurable power is given by the Dicke radiometer equation (multiplied by 2 to take into account the bin spreading, or in fact somewhat less than 2 as can be seen for example in \cite{ADMX})
\begin{equation}
P_{\rm min} = 2 k_B T_c \sqrt{\frac{\Delta f}{t_m}} ,
\label{Pmin1}
\end{equation}
where $t_m$ is the integration time. Similarly the Standard Quantum Limit (SQL) of the detector sensitivity is given by
\begin{equation}
P_{\rm min} = \hbar \omega_a \sqrt{\frac{\Delta f}{t_m}},
\end{equation}
 If we now use the reference values in Equation (\ref{rateaxion}), from Equation (\ref{powerin})  we get a minimum effective magnetic field measurable with a linear amplifier
\begin{equation}
B_a^{\rm linear} = \sqrt{\frac{\hbar \sqrt{\Delta f / t_m}}{\gamma \mu_B n_S   \tau_{\rm min} V_s}} =4.5\times 10^{-21} \,\,
\, {\rm T},
\end{equation}
where we have put $t_m =1.5\times 10^4$ s as in Equation (\ref{ratefinal}). This must be compared with the expected value $B_a =1.8 \times 10^{-22}$~T, thus showing  that a SQL linear amplifier is unsuitable for axion detection, even though its sensitivity is not too far from the required value.

\subsection{Cavity design, magnetic design, directionality}
As can be seen from Equation (\ref{rateaxion}), the expected signal is directly proportional to the total number of spins, i.e. to the volume of the magnetic sample. Working at high frequency requires smaller dimensions for the resonant cavity, and it becomes difficult to fit a consistent amount of material unless special designed cavities are used. The cavity design must also allow for the presence of a uniform magnetic field to magnetize the sample. 
{\color{black}
\begin{figure}[htbp]
\begin{center}
\includegraphics[width=.5\linewidth]{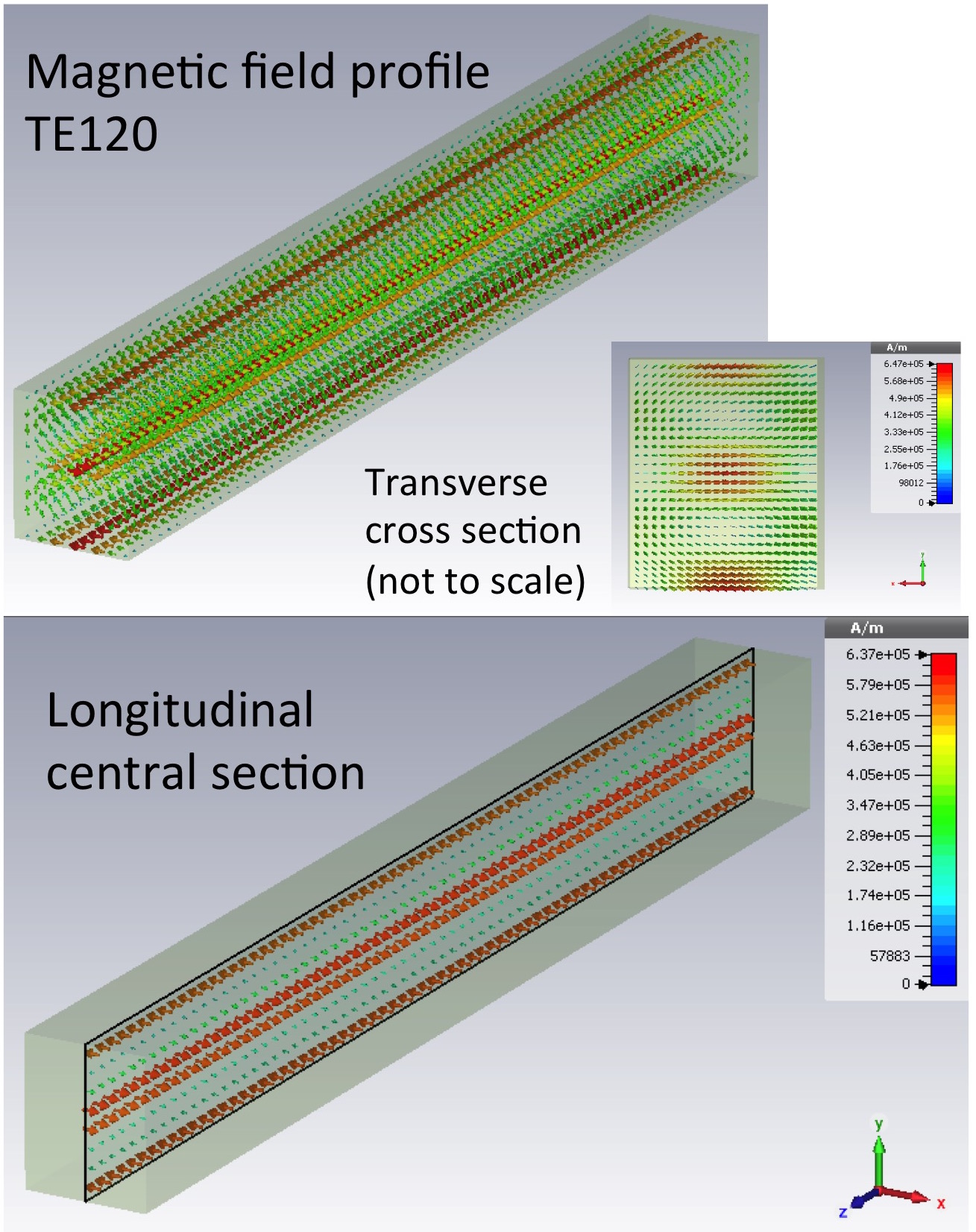}
\caption{Magnetic field profiles of the cavity for the QUAX detector. It is a long single structure with 4 side walls and 2 end caps. The magnetic sample lies at the center and fills the system all along the $z$ direction. The external static magnetic field is solenoidal. The total length (along the $z$ direction) of the system is free since it does not determine the resonance frequency. Its maximum size will be fixed by the size of the magnetic field source and of the refrigerator.}
\label{cavity}
\end{center}
\end{figure}

The QUAX apparatus will use as a base element a rectangular cavity where the $(x,y)$ dimensions determine the resonant frequency of the TE120 mode, i.e. the axion searching frequency $\omega_a$.
The $z$ dimension is free for this mode, and so it can be made as large as possible inside the magnet bore. The static magnetic field $\mathbf{B}_0$ lies in the $z$ direction and  can be easily produced using a long solenoid. 

For this mode TE120, the coefficient $C_{nl}$ appearing in equation (\ref{primakoff}), and related to the rate of the Primakoff effect, is zero, as the rf electric field mode has two lobes with the field parallel to the external magnetic field but opposite directions. Therefore, to look for axion photon coupling, it is  necessary to use other modes, for instance the TE110 for the proposed cavity.

A schematic design for the base cavity is shown in Figure \ref{cavity}. {\color{black}
However, to fulfill the spatial coherence requirement,
the magnetized material must have a dimension smaller than half of the microwave wavelength and axion coherence length  in the $(x,y)$ plane and $z$ direction, respectively.
The necessary volume of 100 cm$^3$ is then obtained by operating 16 microwave base cavities, each containing a rod of magnetized material
of 2 mm diameter and 2 m length. 
All cavities, arranged in a  $4 \times 4$   array in the $(x,y)$ plane, are read by the same microwave counter.  The overall increase of the  thermal photons rate with respect to the base cavity can be compensated  by lowering the
 working temperature from 116 mK to 100 mK. 
 
}}

It is important to stress that with the proposed scheme the static magnetic field is used only to determine the working frequency through Larmor precession and it does  not enter  directly into the signal strength, like for example in the haloscope detector proposed by Sikivie~\cite{Sikivie}. Our approach has the advantage of  reducing the field complexity and amplitude, since  the  typical field is about 2 T.  As a consequence, we can use superconducting cavities  made of NbTi alloy, having a critical magnetic field above 10 T below 4 K \cite{Superconductor}. Such cavities could reach $Q$ values in excess of $10^6$, thus providing the best matching between axion  and detector linewidth. 

\begin{figure} [h]

 \begin{center}
\includegraphics[width=0.7\linewidth]{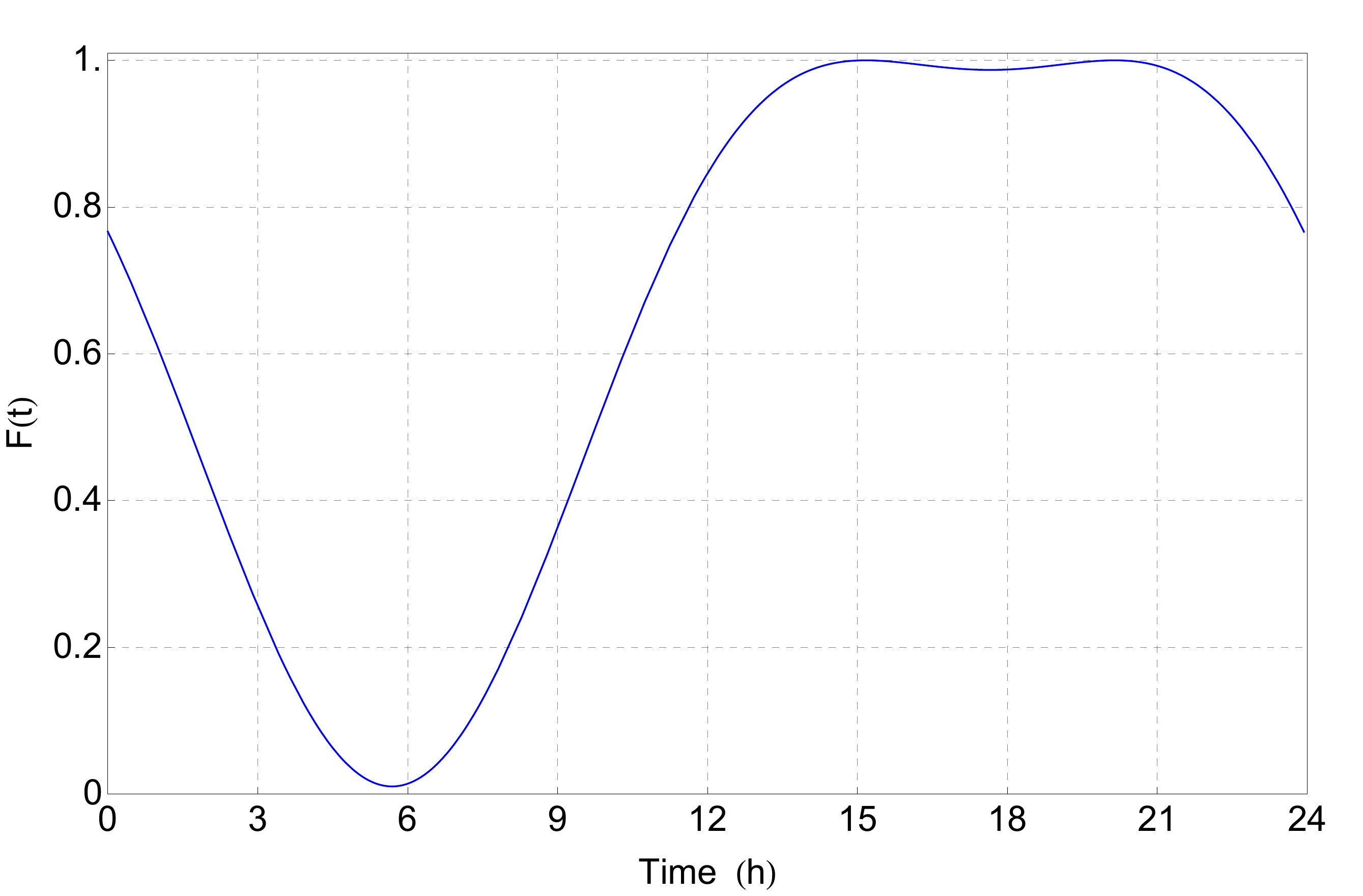}
\caption{Plot of the daily modulation of the axion flux for the QUAX detector located at Legnaro (PD) with 
 a static magnetic field  oriented  North-South and lying in the local horizontal plane:  
 $F(t) \simeq 1  - (0.44+0.072  \cos(0.21 + \omega_\oplus t ) + 
 0.55 \sin(0.21 +  \omega_\oplus t ))^2$, where $\omega_\oplus$ is the Earth rotation rate 
 and $t$ is the universal time coordinate UTC. The time origin is the midnight of northern hemisphere's fall equinox. }
  \label{modulation}
\end{center}

  \end{figure}

The static magnetic field direction must be uniform in all the magnetized sample, and a simple solenoidal design will meet this requirement. Moreover, the presence of the superconductor will help in making the field spatially homogeneous and temporally stable.

The motion of the solar system  with respect to the galactic center leads to a non isotropic 
 axion flux which is peaked in the direction in which the Earth is travelling.
Moreover, due to Earth rotation,  the direction of  the static magnetic field $\mathbf{B}_0\equiv B_0 \mathbf{U}_0$ of 
the QUAX detector changes with time  with respect to the direction of the axion flux.   
The direction $\mathbf{N}_{a}$ of the axion wind  points approximately toward the star Vega of the 
 Cygnus constellation with  equatorial celestial coordinates ($RA = 277.5$ deg, $DEC= 38.8$ deg). 
  Then the daily modulation due to Earth rotation is given by $F=1-(\mathbf{N}_{a}\cdot \mathbf{U}_0)^2$. 
 In Figure \ref{modulation} we have  plotted the directional pattern $F$ for the QUAX detector located at Legnaro (Italy) with 
the static magnetic field  oriented  North-South and lying in the local horizontal plane.  It is worth noticing the strong 
modulation (up to 100 \%) with a sidereal day period of $86164.091$ s that is a strong signature for 
axion dark matter since most of backgrounds are not expected to show this kind of time dependence. 
 
 {\color{black}
 We have shown that we can also search for photons produced via Primakoff effect 
 by tuning the photon detector to the TE110 mode of the cavity.
As only photons produced via the Primakoff effect and thermal photons are stored in this cavity mode, their rate is insensitive to the orientation of the static magnetic field with respect to the direction of the axion flux.  
The two measurements scheme are  complementary: QUAX could be operated in the axion-electron coupling mode when the directionality allows for maximum signal, and in the axion-photon coupling mode in the remaining time spans.

If a signal were found either in QUAX or in any other experiment at a frequency $\overline{\omega}_a$ \cite{Rosemberg}, the operation of QUAX in the TE120 mode of the cavity and with the static magnetic field oriented as indicated in Fig. \ref{cavity} would be an optimal way to confirm possibly the axion origin of the signal itself.

 }

\section{A preliminary test}

We have tried to verify experimentally the correspondence between the thermal noise in an empty cavity or in a cavity filled with a magnetic sample. Since  quantum counter devices are not yet available, we have measured with a linear amplifier the power delivered by a microwave resonant cavity in three different conditions: a) an empty cavity; b) a cavity with a magnetic material inside but no magnetizing field; c) with a magnetic material subjected to a magnetizing field such that hybridization occurs. We have carefully checked that  measurements a) and b) give the same results (apart from a small frequency shift),  and so we report only on the comparison between b) and c) conditions. 
\begin{figure}[htbp]
\begin{center}
\includegraphics[width=.7\linewidth]{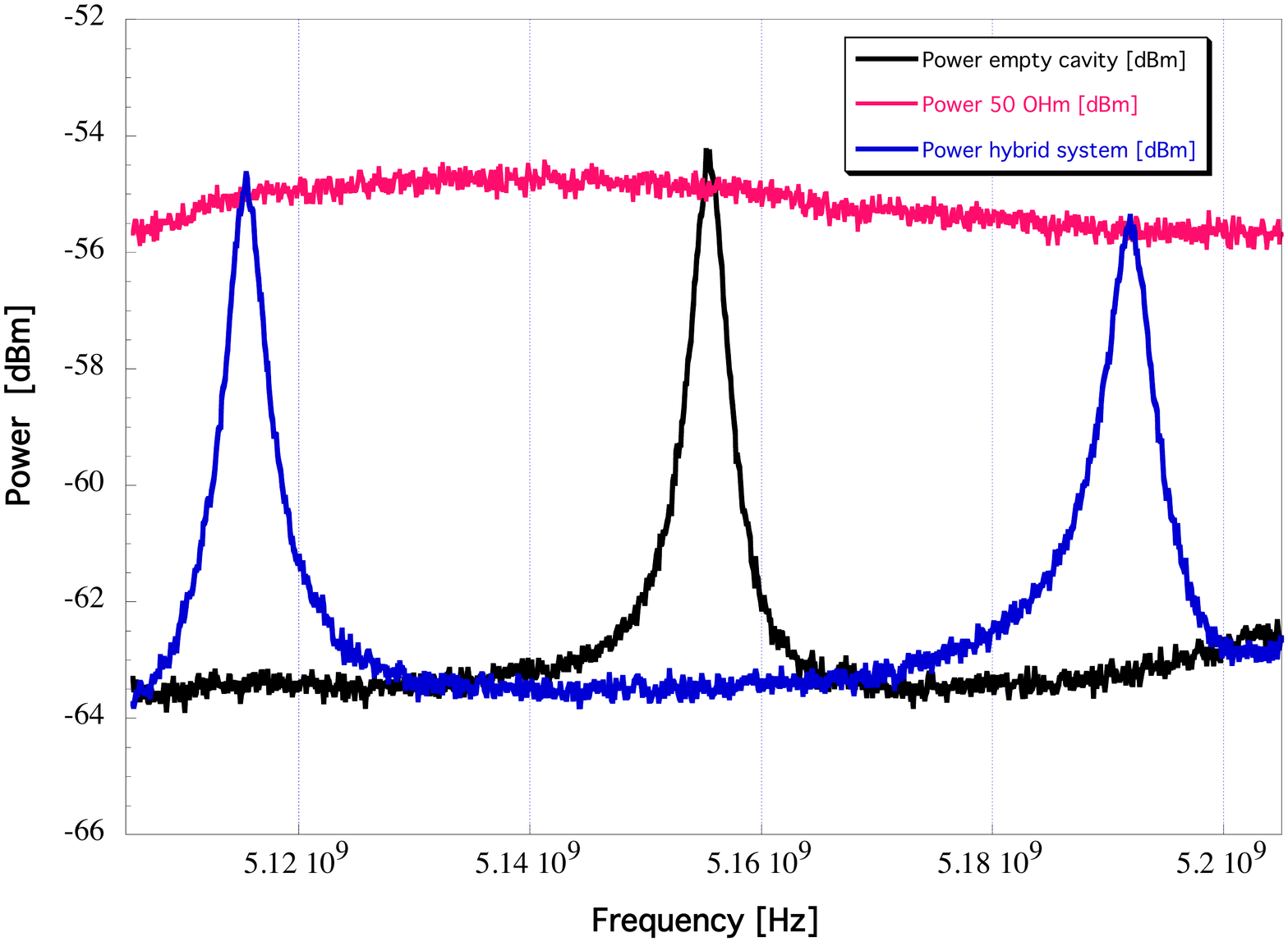}
\caption{Thermal noise measured with a linear amplifier (total gain = 69 dB, RBW = 100 kHz). $T_c = 300$ K. The antenna coupling has been optimized for each individual measurement. Red curve: amplifier input connected to a 50 $\Omega$ load. Black curve: empty cavity output. Blue curve: hybrid system output.}
\label{noise}
\end{center}
\end{figure}

The microwave resonator, used for this test, is a copper cavity shaped in the form of a parallelepiped with dimensions along $x,y,z$ axes 85 mm, 40 mm and 10 mm, respectively.
 The  resonance frequency of the selected TE102 mode is $\omega_c / 2 \pi = 5.154$ GHz, with a total cavity linewidth  $k_c/2\pi=2.4 $ MHz. The magnetic material is a sphere of YIG (Yttrium Iron Garnet) ($n_S = 2 \times 10^{28}$ m$^{-3}$) of 2 mm diameter placed exactly at the center of the cavity. 
With a magnetizing field $B_0= 0.19 $ T, placed along the $z$ direction of the cavity, i.e. orthogonal to the radio frequency magnetic field, the system hybridizes. A mobile loop antenna is critically coupled to the cavity mode, and measure the power delivered by the cavity also in the absence of any input, the residual emission being due to the thermal photons stored in the resonator. The antenna output is amplified by a cascade of two cryogenic amplifiers (kept at a temperature of 77 K), the first one having a noise temperature well below the cavity temperature. The final output, amplified by about 69 dB, 
is then fed into a spectrum analyzer for recording.
The results of these measurements are given in Figure \ref{noise}. For calibration purposes a 50 Ohm load was put at the input of the amplifying chain: this allows to check the gain versus frequency curve and the nominal input level. As can be seen from the figure, the cavity peak power level correspond to the Johnson (Thermal) noise of the 50 Ohm resistor. 

When the system is hybridized, two peaks appear with a separation of about 80 MHz. The expected value is $g_m/2\pi = 90 $ MHz. From the fit of the data with Equation (\ref{hyb}) one gets $k_m/2 \pi=1.0 $ MHz, which corresponds to the natural linewidth of YIG associated with the spin-spin relaxation time $t_2 = 0.16\, \mu$s.
It is evident that the height of the two hybridized modes coincides again with the Johnson noise of the 50 Ohm load and consequently of the empty cavity. Checking this fact at different temperatures will allow us to infer that thermal noise in a cavity filled with a magnetized sample is the same as in an empty cavity. 

To give an example of the effectiveness   of our approach, we can compute the sensitivity corresponding to the measurement of figure \ref{noise}. 
The best choice for the axion detection is to use as measuring bandwidth the axion one: since the central frequency is 5.1 GHz, 
the corresponding measurement bandwidth is $\Delta f=2.5$ kHz. The Johnson noise power delivered by the cavity at its center frequency is
\begin{equation}
P_{\rm Johnson} = k_B T_c \Delta f = 1.0 \times 10^{-17} \,\,{\rm W}.
\end{equation}
By inverting Equation (\ref{powerin}), we can calculate the equivalent axion field giving the same power signal at the cavity output
\begin{equation}
\label{bsens}
B_a^{\rm sens}=\sqrt{\frac{P_{\rm Johnson}}{\gamma \mu_B n_S  \omega_a   \tau_{\rm min} V_s}}.
\end{equation}
Putting $\omega_a/2 \pi = 5.1$ GHz ($m_a = 21 \,\mu$eV), $\tau_{\rm min}=0.08\, \mu$s and 
$V_s = 18$ mm$^3$, we have
\begin{equation}
B_a^{\rm sens}=1.4 \times 10^{-14} \,\, {\rm T},
\label{roomt}
\end{equation}
which is about $10^9$ times larger than the expected value of $B_a = 2.0 \times 10^{-23} $ T for this axion mass. In the experimental test described above, using Eq. (\ref{Pmin1}) with an integrated time of one hour, the minimum measurable effective magnetic field is $B_{\rm min}=4\times 10^{-16} $ T. For a SQL device the corresponding limit in one hour of measurement is $7\times 10^{-18}$ T. While these values are far from the needed sensitivity, 
we have to remember that this is a room temperature device with a very small magnetic sample. 

\section{Conclusions}

We have presented a novel experimental scheme to detect galactic axions through their coupling to the electron spin in a magnetized media.
 The use of the Larmor resonance of a material allows enough sensitivity to search for the axion, in the DFSZ model, as a primary component of the local dark matter density. {\color{black}By using a suitable cavity mode we can also  exploit the axion photon coupling via the Primakoff effect.}  
The crucial issue is the concurring development of a single photon counter in the microwave regime. The possibility of working in an ultra-cryogenic environment might allow the scanning of an interesting mass range expected for the QCD axion.

\section*{Acknowledgments}

The authors wish to thank Giuseppe La Rocca for useful discussion concerning the radiation damping and  Enrico Berto for mechanical working on the preliminary set-up.

\section*{References}





\end{document}